# Publication modalities 'article in press' and 'open access' in relation to journal average citation

Sara M. González-Betancor *

Universidad de Las Palmas de Gran Canaria, Departamento de Métodos Cuantitativos en Economía y Gestión, Campus de Tafira, 35017 Las Palmas de Gran Canaria, Spain. *E-mail*: sara.gonzalez@ulpgc.es

ORCID: http://orcid.org/0000-0002-2209-1922

Pablo Dorta-González

Universidad de Las Palmas de Gran Canaria, TiDES Research Institute, Campus de Tafira, 35017 Las Palmas de Gran Canaria, Spain. *E-mail*: pablo.dorta@ulpgc.es

ORCID: http://orcid.org/0000-0003-0494-2903

* Corresponding author

# Publication modalities 'article in press' and 'open access' in relation to journal average citation

**Abstract:**

There has been a generalization in the use of two publication practices by scientific journals during the past decade: 1. 'article in press' or early view, which allows access to the accepted paper before its formal publication in an issue; 2. 'open access', which allows readers to obtain it freely and free of charge. This paper studies the influence of both publication modalities on the average impact of the journal and its evolution over time. It tries to identify the separate effect of access on citation into two major parts: early view and selection effect, managing to provide some evidence of the positive effect of both.

Scopus is used as the database and CiteScore as the measure of journal impact. The prevalence of both publication modalities is quantified. Differences in the average impact factor of group of journals, according to their publication modalities, are tested. The evolution over time of the citation influence, from 2011 to 2016, is also analysed. Finally, a linear regression to explain the correlation of these publication practices with the CiteScore in 2016, in a *ceteris paribus* context, is estimated.

Our main findings show evidence of a positive correlation between average journal impact and advancing the publication of accepted articles, moreover this correlation increases over time. The open access modality, in a *ceteris paribus* context, also correlates positively with average journal impact.

## 1. Introduction

The communication of research findings has benefited greatly from the emergence of the Internet, and especially from two publication practices during the past decade. The first practice is the publication of documents under the *Article in Press* (AIP) modality, also known as online first. The AIP practice allows access to the documents from the moment the version is accepted by the editor, and before its formal publication in an issue. The second practice is publication under the *Open Access* (OA) modality, allowing readers freely available access to the documents. With these publication modalities publishers and authors aim to increase the visibility, usage, and citation impact of the document. However, to date no strong evidence has been found as to whether such modalities do in fact have the desired effect.

Open Access was defined in 2002 by *Budapest Open Access Initiative* as free and unrestricted access on the public Internet to literature that scholars provide without expectation of direct payment (Prosser, 2003). Open access is not limited to just two modalities, though the gold and green are the main roads. *Gold OA* refers to articles in fully accessible OA journals, and *green OA* refers to publishing in a traditional journal in addition to self-archiving the preprint or postprint paper in a repository (Harnad et al., 2004). With respect to gold OA, the *Directory of Open Access Journals* (DOAJ) is currently the largest index presenting quality controls. According to the DOAJ list, in April 2019 there were 12,980 OA journals, of which 9,507 were totally free and 3,415 had article processing charges (APC). There was no available information about the possible existence of APC for 58 journals.

The OA research has produced a significant body of literature. For recent and detailed reviews, see Tennant et al. (2016) and McKiernan et al. (2016). And for a definition of OA and its subtypes, to assess the prevalence of OA and to examine the relative impact of OA citations, see Piwowar et al. (2018). Here, instead, we briefly review the literature on the OA citation advantage.

Some researchers, starting with Lawrence (2001), have found that OA articles tend to have more citations than pay-for-access articles. This OA citation advantage has been observed in a variety of academic fields including computer science (Lawrence, 2001), philosophy, political science, electrical & electronic engineering, and mathematics (Antelman, 2004), physics (Harnad et al., 2004), biology and chemistry (Eysenbach, 2006), civil engineering (Koler-Povh, Južnič & Turk, 2014), as well as biology and medicine (Li et al., 2018).

However, there is no general agreement in the literature about the OA citation advantage (Davis et al., 2008; Norris, Oppenheim & Rowland, 2008; Joint, 2009; Gargouri et al., 2010; Wang et al., 2015; Dorta-González, González-Betancor & Dorta-González, 2017; Dorta-González & Santana-Jiménez, 2018). Furthermore, some authors are critical of the causal link between OA and higher citations, stating that the benefits of OA are uncertain and vary among different fields (Craig et al., 2007; Davis & Walters, 2011).

Kurtz et al. (2005), and subsequently other authors (Craig et al., 2007; Moed, 2007; Davis et al., 2008), set out three postulates supporting the existence of a correlation between OA and increased citations: *(i)* OA articles are easier to obtain; and therefore easier to read and cite (*Open Access postulate*); *(ii)* OA articles tend to be available online prior to their publication and therefore begin accumulating citations earlier than pay-for-access articles (*Early View postulate*); and *(iii)* more prominent authors are more likely to provide OA to their articles, and authors are more likely to provide OA to their highest quality articles (*Selection Bias postulate*). Moreover, these authors conclude that early view and selection bias effects are the main factors behind this correlation.

Gaule & Maystre (2011) and Niyazov et al. (2016) found evidence of *Selection Bias* in OA, but still estimated a statistically significant citation advantage even after controlling for that bias. However, Björk and Solomon (2012) argued that the distribution model is not related to journal impact. This result was confirmed by Solomon, Laakso & Björk (2013), who concluded that articles are cited at a similar rate regardless of the distribution model.

The *Early View* postulate is also related to the publication of AIPs and self-archiving by authors. Nowadays, many journals post accepted articles online before they are formally published in an issue (in-press articles). The overall publication delay (the time between submission and publication) negatively influences citations (Luwel & Moed, 1998; Yu, Wang & Yu, 2005; Tort, Targino & Amaral, 2012; Shi, Rousseau, Yang & Li, 2017). Conversely, advance the publication of in-press articles increases citations (Alves-Silva et al., 2016; Al & Soydal, 2017; Echeverría, Stuart, & Cordón-García, 2017). Thus, many publishers provide access to in-press articles to minimize the effect of publication delays and potentially increase citations. This is the case, among others, of Elsevier (articles in press), Nature Publishing Group (advance online publication), Springer (online first), Taylor & Francis (latest articles), and Wiley (early view).

Some authors archive the preprint or postprint (final draft after peer review) of their articles to OA repositories or share them via social networks before they are available on the publisher's website. This strategy reduces the effect of publication delays and makes it more likely that an article will be cited before it is formally published. According to the *SHERPA/RoMEO* database of publishers' policies on copyright and self-archiving, 81% of publishers formally allow some form of self-archiving (statistics for the 2,561 publishers in the database; accessed April, 2019).

Therefore, open access and in-press access are two increasingly important phenomena that need to be investigated, especially in terms of any possible interrelationships between the two modalities. One of the limitations seen in the related literature, whether supporting or refuting the citation advantage, is often the small number of analysed articles. Another limitation concerns the concentration of a study on a particular scientific area, especially given the well-known existence of important differences between areas in terms of publication and citation habits (Dorta-González & Dorta-González, 2013; Dorta-González, Dorta-González & Suárez-Vega, 2015). In view of all the above, in this paper we conduct a large-scale analysis of the citation advantage in both publication modalities.

*Objectives and relationship between OA publishing and AIP publishing*

OA publishing and AIP publishing have previously been explored in the literature around strategies that journals can employ to increase the impact factor/citations metrics of a journal. In an environment where publishing is slow and time for various processes are counted in months and sometimes years, AIP publishing can be a good strategy to compete in time with OA publishers, which are generally more efficient in time. Many publishers now have their own megajournals (i.e., Plos One, Scientific Reports, Nature Communications, Science Advances, SAGE Open,...). These megajournals publish continuously so they do not have queues of documents to be published, significantly reducing the times of publication.

The publication delay traditionally refers to the time between the acceptance of an article and its publication and indexing in scientific databases. This delay has been previously proposed to correlate negatively with journal impact factors (Yu, Wang & Yu, 2005). With the development of online access, however, AIP publishing has become commonplace. Thus, a significant fraction of publication delay now consists of a period in which an article is available online, but has not been formally published in print. AIP publishing increases the impact factor of a journal (Tort, Targino & Amaral, 2012). This is because the AIP allows journals to formally publish papers which are 'born' already with citations.

The research question about the difference in citations of open access (as published by full OA journals) to citations in subscription journals has been covered quite extensively (most recently by Li et al., 2018; Dorta-González & Santana-Jiménez, 2018; and Dorta-González, González-Betancor & Dorta-González, 2017). Accordingly, the objective of this study is focusing on the influence of early view availability to average journal citations, and whatever synergies that can create with the open access research question.

Therefore, we address the following questions: *(i)* is there evidence to confirm that advancing the publication of in-press articles improves the average impact of the journal? *(ii)* is there evidence to confirm that open access improves the average journal impact in a *ceteris paribus* context?

## 2. Methodology

*The data*

Part of the final dataset was downloaded directly from the Scopus website at https://journalmetrics.scopus.com, with the rest obtained from Scopus by request.

The first file was downloaded in summer 2017 and contains information of all journals included in Scopus for a 6-year period, starting in 2011. We restricted our analyses to journals with CiteScore in 2016, amounting to a total of 21,529 journal titles. Each journal is classified according to a system of 329 subject areas. The downloaded dataset includes the following variables of interest for our research for each year from 2011 to 2016:

- Scopus ID: a unique identifier for every journal.
- CiteScore: this is a measure of average journal impact calculated by Scopus. This indicator measures the average citations received per document published in the journal in a window of three years (e.g. CiteScore 2016 is obtained by dividing the number of citations in 2016 to all documents published from 2013 to 2015 by the number of documents published from 2013 to 2015).
- Open Access (OA): journals covered by Scopus are catalogued as OA if listed in the DOAJ or the *Directory of Open Access Scholarly Resources* (ROAD).

- The ISSN as well as the e-ISSN are also given.

Scopus also offers other journal metrics in addition to CiteScore, like SJR and SNIP, which are among the set of citation metrics offered by Scopus since 2010. However, the recent CiteScore is a much simpler citation metric -in contrast to SJR and SNIP- that can be easily validated. In this respect, we decided to use CiteScore, instead of a different journal metric, because of its transparency and simplicity to users.

The second file, Scopus Source List, was updated to April 2017 and obtained by request. It has one entry per title with information related to each publisher:

- Scopus ID (used to merge both files).
- Articles in Press (AIP): indicating whether the journal publishes accepted articles before its official publication in an issue, and if they are considered when computing the CiteScore.
- Coverage: indicating the years that the journal has been indexed in Scopus.
- Article language: indicating the different languages in which the title publishes its articles.
- Branch: indicating the scientific branch to which the title belongs by grouping the Scopus classification system into 5 main branches. Titles with more than one subject category can be assigned to more than one branch.
- Publisher's name.
- Publisher's country.

*New variables*

In the first file we generated the variable 'Number of subject categories', describing the number of subject categories to which each title is assigned (out of 329). This variable ranges from one to thirteen and is asymmetrically distributed with a median of two.

In the second file we generated the variable 'Number of branches'. This variable ranges from one to four and is asymmetrically distributed with a median of one.

We also generated a variable called 'Number of languages', as some journals publish their articles in different languages while others publish all their papers in just one. The maximum number of editing languages of these journals is six.

We also recoded the variable 'Publisher's Name', given the excessively large number of categories that the original variable had (11,387). First, we generated a new variable of number of journals per publisher. This variable had a wide range, showing publishers with just one title and others with more than 2,000. Thus, we decided to recode the original variable into a different one called 'Publisher', with one category for each well-known publisher (Elsevier, Emerald, SAGE, Springer Nature, Taylor & Francis, and Wiley-Blackwell) -all with more than 350 titles- and all other publishers -all with less than 350 titles- grouped into a single category called 'Others'.

We also decided to recode the variable 'Publisher's Country' for the same reason. We grouped all countries into continents excluding the four major publishing countries which were given their own category: Germany, Netherlands, United Kingdom, and the United States.

Finally, we decided to differentiate certain characteristics for all publishers. We generated one variable showing the percentage of OA journals for each publisher, and another showing the percentage of journals that publish in-press articles for each publisher.

*Merged dataset*

After generating all the variables we merged both datasets. The final combined dataset includes all variables that could explain -and be correlated with- the journal average impact, measured in terms of its CiteScore:

- Articles in Press (AIP): dichotomous variable that indicates if the journal publishes articles in press or not (0 = No; 1 = Yes).
- Open Access (OA): dichotomous variable that indicates if it is an OA journal or not, i.e., registered in DOAJ and/or ROAD by April 2017 (0 = No; 1 = Yes).
- e-ISSN: dichotomous variable that indicates if the journal is accessible on-line or not (0 = No; 1 = Yes).
- Indexed year: discrete variable that indicates the first year the journal was indexed in the Scopus database.
- Number of languages: discrete variable for the number of editing languages in the journal.
- English/Spanish/Chinese: dichotomous variables that indicate if the journal articles are edited in English/Spanish/Chinese or not (0 = No; 1 = Yes).

- Number of subject categories/branches: discrete variable that describes the number of subject categories/branches the journal is included under.
- Branch: categorical variable that indicates the branch in which the journal is included (Health Sciences, Life Sciences, Physical Sciences, Social Sciences, General). Journals assigned to more than one branch were included under the category "Variety of branches".

The final database also contains variables related to the journal's publisher:

- Country: categorical variable that indicates the publisher's geographical area (Africa, America, Asia, Europe, Oceania, Germany, Netherlands, United Kingdom, and the United States).
- Publisher: categorical variable that indicates the name of the publisher (Elsevier, Emerald, SAGE, Springer Nature, Taylor & Francis, Wiley-Blackwell, Others).
- Percentage of AIP/OA journals: continuous variable that indicates the amount of AIP/OA journals in relation to the total amount of journals of each publisher.

## 3. Results and Discussion

In this section we quantify both publication modalities (AIP and OA) and perform a bivariate analysis to test hypotheses on the relationship between the CiteScore and the AIP and OA variables, differentiating among groups by geographical area or language. Finally, we estimate a linear regression to explain the correlation between all considered variables and the CiteScore 2016, in a *ceteris paribus* context.

*Prevalence of Articles in Press and Open Access*

The prevalence of the AIP and gold OA publication modalities was quantified (Table 1). One out of every two journals (10,475) indexed in the Scopus database in 2016 used one or other of these publication modalities. To be more precise, 37% (7,953) of all journals published under the AIP practice, while 16% (3,522) of all journals used the gold OA modality. Only 2.3% (500) of all journals used both publication practices.

*Table 1. Prevalence of publication modalities AIP and OA in number of journals 2016*

|  |  | Articles in Press |  |  |
|---|---|---:|---:|---:|
|  |  | No | Yes | Total |
| Open Access | No | 10,554 | 7,453 | 18,007 |
|  | Yes | 3,022 | 500 | 3,522 |
|  | Total | 13,576 | 7,953 | 21,529 |

Source: Scopus

The CiteScore difference of journals according to whether or not they publish Articles in Press is statistically significant, showing a great advantage -more than double- in favour of journals that do publish AIP (Table 2). On the contrary, the CiteScore difference of journals according to whether or not they are OA journals shows a slight statistically significant advantage in favour of journals that do not use the gold OA modality. Both differences are statistically significant, although the former is considerably more pronounced than the latter. Thus, it seems that open access journals (gold OA) have lower CiteScores than non-OA journals (Table 2, second column), and publishing AIP means having higher CiteScores than not publishing them (Table 2, first column).

*Table 2. Average CiteScore 2016 by publication modalities*

|  | Articles in Press | Open Access |
|---|---:|---:|
| No | 0.872 | 1.298 |
| Yes | 1.965 | 1.164 |

Source: Scopus

However, considering both types of publishing at the same time, the relation between gold OA and CiteScore reverses. Without taking into account the AIP variable, non-OA journals show to have a higher journal impact (Table 2). But, by splitting the sample of journals in two groups -those that do and those that do not publish AIP-, OA journals show to have higher impact factor in both groups (Figure 1).

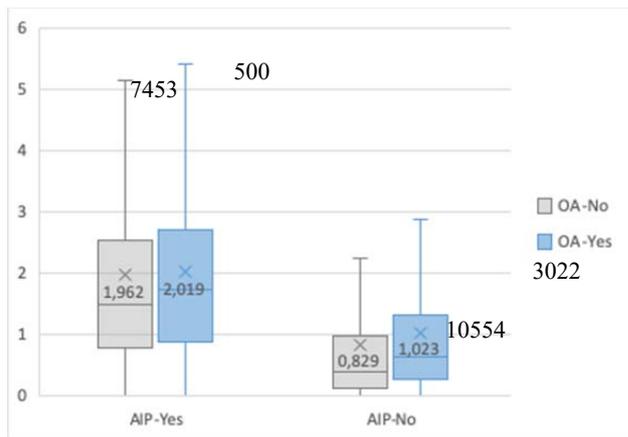

*Figure 1. Distribution of CiteScore 2016 by type of publication modalities (AIP and OA) with its mean. Number of journals at the top of each distribution. Outliers are not shown, to keep the Y-axis on a scale that facilitates comparison between distributions.*

This could seem to be a paradox, yet it is just another example of the so called Simpson's paradox or Yule-Simpson effect (Yule, 1903; Simpson, 1951), which states that a trend appears in several different groups of data but disappears or reverses when these groups are combined. Although it could seem to be counter-intuitive, the paradox in this case is easily explained by the total frequency of each type of journals in each group. In the OA group of journals, very few of them publish AIP (500 out of 3,522), which are the journals with higher impact (Figure 1). And the opposite happens in the group of non-OA journals, where the proportion of journals that publish AIP is much higher (7,453 out of 18,007), rising the journal impact of the whole group of non-OA journals.

As can be seen in Figure 1, within the non-OA group, journals that use the in-press modality have an average impact which is more than double that of journals which do not. Likewise, within the OA group, the average impact of journals that use the in-press modality is almost twice as high as that of journals that do not. Thus it can be concluded that there is evidence that advance publication of accepted articles is the most important factor in terms of citation advantage, even though this type of journals have a greater variability.

*Bivariate analysis*

For a more in-depth analysis of the relation of the publication modalities on the journal average citation, we tested for statistically significant differences in the CiteScore over the last six years depending on the two publication practices analysed.

*Table 3. Two-sample t-tests for equality of means (OA vs. non-OA journal)*

|  | Open Access | N | Mean | S.D. | p-value | 95% CI for the diff. | |
|---|---|---|---|---|---|---|---|
| CiteScore 2011 | No | 16,077 | 1.239 | 1.989 | 0.000 | 0.185 | 0.315 |
|  | Yes (12.2%) | 2,237 | 0.989 | 1.378 |  |  |  |
| CiteScore 2012 | No | 16,940 | 1.243 | 2.030 | 0.000 | 0.208 | 0.322 |
|  | Yes (13.6%) | 2,675 | 0.978 | 1.264 |  |  |  |
| CiteScore 2013 | No | 17,657 | 1.277 | 2.083 | 0.000 | 0.178 | 0.290 |
|  | Yes (14.4%) | 2,979 | 1.043 | 1.300 |  |  |  |
| CiteScore 2014 | No | 18,176 | 1.231 | 2.006 | 0.000 | 0.118 | 0.229 |
|  | Yes (14.8%) | 3,161 | 1.058 | 1.367 |  |  |  |
| CiteScore 2015 | No | 18,674 | 1.259 | 1.964 | 0.000 | 0.073 | 0.185 |
|  | Yes (15.3%) | 3,369 | 1.130 | 1.444 |  |  |  |
| CiteScore 2016 | No | 18,874 | 1.283 | 1.999 | 0.000 | 0.066 | 0.175 |
|  | Yes (15.8%) | 3,535 | 1.163 | 1.399 |  |  |  |

The number of OA journals in Scopus experienced a slight but regular increase over the 6-year period, representing 12.2% of the whole dataset in 2011 and 15.8% in 2016 (Table 3). The mean CiteScore of OA journals is slightly lower than the mean CiteScore of non-OA journals. This difference is statistically significant in every year, but decreases with time, ranging between 0.185 and 0.315 in 2011 and decreasing to a range between 0.066 and 0.175 in 2016. Thus, although the average impact of OA journals is slightly lower than that of non-OA (Dorta-González et al., 2017), empirical evidence indicates that this difference is gradually becoming smaller and tending to disappear.

The proportion of journals that publish AIPs (e.g. 35.9% in 2016, Table 4) is higher than the proportion of OA journals (e.g. 15.8% in 2016, Table 3) in the whole dataset. Table 4 shows that there is also a statistically significant difference in CiteScore in favour of those journals that publish articles in press compared to those that do not. This difference (AIP vs non-AIP) is much larger than that of Table 3 (OA vs. NOA) and has been increasing over the last six years, ranging from -1.000 to -0.888 in 2011 and from -1.145 to -1.036 in 2016. This reveals evidence in favour of the publication of AIPs.

Moreover, the citation advantage to those journals that use this publication modality increases with time.

*Table 4. Two-sample t-tests for equality of means (AIP vs. non-AIP)*

|  | Article in Press | N | Mean | p-value | 95% CI for the diff. | |
|---|---|---|---|---|---|---|
| CiteScore 2011 | No | 11,136 | 0.838 | 0.000 | -1.000 | -0.888 |
|  | Yes (39.2%) | 7,178 | 1.782 |  |  |  |
| CiteScore 2012 | No | 12,235 | 0.834 | 0.000 | -1.048 | -0.936 |
|  | Yes (37.6%) | 7,380 | 1.826 |  |  |  |
| CiteScore 2013 | No | 13,076 | 0.855 | 0.000 | -1.118 | -1.002 |
|  | Yes (36.6%) | 7,560 | 1.915 |  |  |  |
| CiteScore 2014 | No | 13,594 | 0.829 | 0.000 | -1.093 | -0.982 |
|  | Yes (36.3%) | 7,743 | 1.867 |  |  |  |
| CiteScore 2015 | No | 14,135 | 0.863 | 0.000 | -1.102 | -0.996 |
|  | Yes (35.9%) | 7,908 | 1.912 |  |  |  |
| CiteScore 2016 | No | 14,362 | 0.873 | 0.000 | -1.145 | -1.036 |
|  | Yes (35.9%) | 8,047 | 1.963 |  |  |  |

We also graphically analysed whether there were differences in the CiteScore depending on geographical area and the publication (or not) of AIPs (Figure 2). In this case, we decided to employ another central position measure, the median CiteScore instead of the mean, in case of any doubts about the accuracy of the latter.

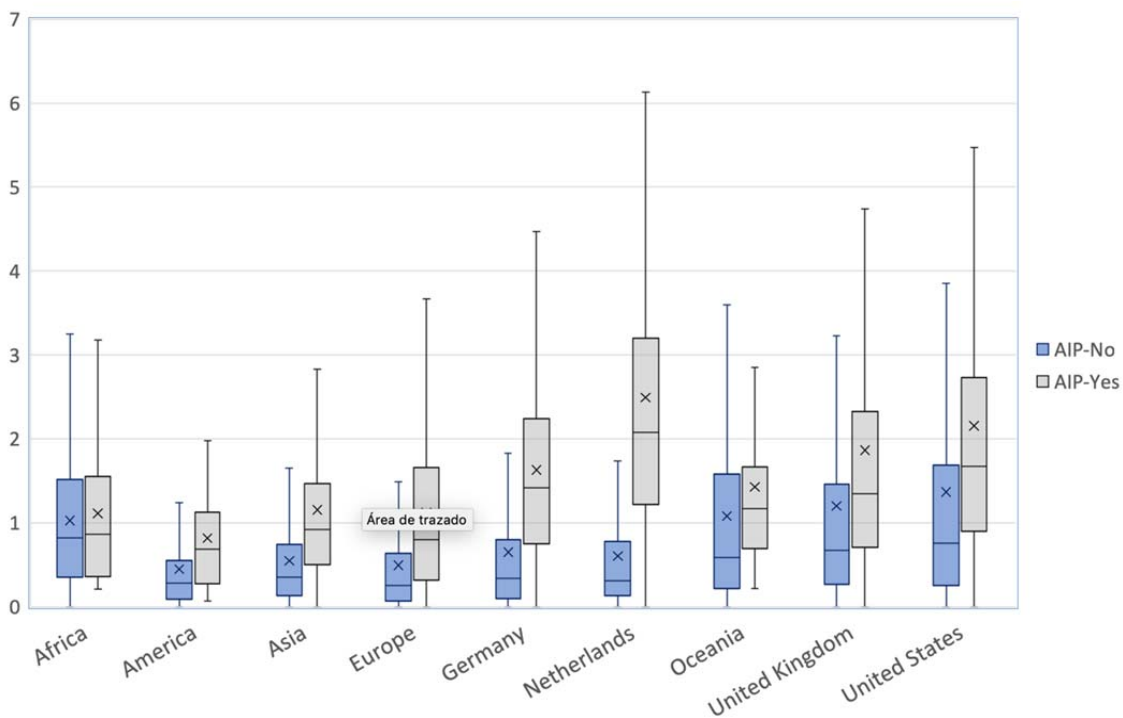

*Figure 2. Distribution of CiteScore 2016 by geographical area and publication of AIP. America refers to the rest of the continent excluding the United States, and Europe refers to the rest of the continent excluding Germany, the Netherlands, and the United Kingdom. Outliers are not shown, to keep the Y-axis on a scale that facilitates comparison between distributions.*

In almost all the geographical areas the pattern remains the same, with the CiteScore of journals that publish AIPs higher than the CiteScore of those that do not. With the exception of Africa, there is a marked difference for all geographical areas.

Table 4 shows that publishing AIPs is related to having a higher CiteScore, though this could be because the journals that publish AIPs are more likely to publish in English as the universal language of science. Thus, we decided to test if the difference in CiteScore, apart from publishing AIPs, could be related to the language of publication. We therefore tested for differences in the CiteScore of journals that publish articles in English compared to those that do not publish in that language, and according to whether the journal publishes AIPs or not.

*Table 5. Two-sample t-tests for equality of means of CiteScore 2016 (Articles in English vs. Articles not in English)*

| Articles in Press | English | N | Mean | p-value | 95% CI for the diff. | |
|---|---|---|---|---|---|---|
| No | No | 2,072 | 0.248 | 0.000 | -0.801 | -0.729 |
|  | Yes | 11,490 | 1.013 |  |  |  |
| Yes | No | 285 | 0.421 | 0.000 | -1.676 | -1.529 |
|  | Yes | 7,695 | 2.024 |  |  |  |

The number of journals that publish articles in English is clearly higher in both groups (Table 5), accounting for 85% of the group of journals that do not publish AIPs and 96% of the group that do. Table 5 shows a clear statistically significant difference in CiteScore in favour of journals that publish articles in this language in both groups. This difference is around 0.7 to 0.8 for the first group (AIP = No), and around 1.5 to 1.7 for the second group (AIP = Yes). This is strong evidence of something that is well known: research that is communicated in English has more impact.

*Linear regression*

All conclusions that are derived from a bivariate analysis have to be confirmed through a more robust estimation technique. Thus, in this section we estimate a linear regression for the CiteScore 2016, the most updated value that we have, explained by the variables of interest, and all other available variables that could also influence on the CiteScore and that therefore have to be included as control variables. The results of this estimation are shown in Table 6.

At the beginning of the table we show the variables of interest, some of which are related to the journals and others to their publishers. After this group of variables, we show all control variables, with some also related to the journals and some to their publishers.

*Table 6. Ordinary Least Squares estimations for CiteScore 2016*

| Variables | Coef. | t | Sig. | Beta | Beta CI 95% |
|---|---|---|---|---|---|
| Constant | 20.503 | 14.37 | *** | | |
| **Variables of interest** | | | | | |
| *Related to journals* | | | | | |
| Open Access (Ref. No) | 0.265 | 4.80 | *** | 0.054 | (0.037 / 0.071) |
| Articles in Press (Ref. No) | 0.537 | 11.62 | *** | 0.145 | (0.125 / 0.165) |
| *Related to publishers* | | | | | |
| Percentage of OA journals | 0.001 | 1.74 | * | 0.020 | (-0.011 / 0.051) |
| Percentage of AIP journals | -0.001 | -1.54 | | -0.023 | (-0.054 / 0.008) |
| **Control variables** | | | | | |
| *Related to journals* | | | | | |
| e-ISSN (Ref. Only print) | 0.079 | 2.51 | ** | 0.021 | (0.006 / 0.036) |
| Indexed year | -0.011 | -14.96 | *** | -0.100 | (-0.119 / -0.081) |
| Number of languages | -0.153 | -4.82 | *** | -0.035 | (-0.049 / -0.021) |
| English (Ref. No) | 0.476 | 10.16 | *** | 0.084 | (0.068 / 0.100) |
| Spanish (Ref. No) | 0.002 | 0.02 | | 0.000 | (-0.014 / 0.014) |
| Chinese (Ref. No) | 0.247 | 2.53 | ** | 0.018 | (0.005 / 0.031) |
| Number of subject categories | 0.034 | 3.18 | *** | 0.023 | (0.009 / 0.037) |
| Number of branches | 0.789 | 18.40 | *** | 0.203 | (0.184 / 0.222) |
| Branch (Ref. Variety of branches) | | | | | |
|   Health Sciences | 0.899 | 17.40 | *** | 0.219 | (0.193 / 0.245) |
|   Life Sciences | 1.091 | 14.71 | *** | 0.133 | (0.118 / 0.148) |
|   Physical Sciences | 1.226 | 21.10 | *** | 0.290 | (0.263 / 0.317) |
|   Social Sciences | 0.492 | 8.52 | *** | 0.126 | (0.101 / 0.151) |

| | | | | |
|---|---:|---:|---|---:|---|
| General | 1.340 | 6.45 | *** | 0.042 | (0.030 / 0.054) |
| *Related to publishers* | | | | | |
| Percentage of e-ISSN journals | 0.002 | 3.93 | *** | 0.036 | (0.003 / 0.069) |
| Country (Ref. United States) | | | | | |
|   Africa | -0.776 | -7.30 | *** | -0.049 | (-0.061 / -0.037) |
|   America | -0.713 | -10.24 | *** | -0.082 | (-0.096 / -0.068) |
|   Asia | -0.857 | -18.76 | *** | -0.154 | (-0.169 / -0.139) |
|   Europe | -0.700 | -15.88 | *** | -0.140 | (-0.157 / -0.123) |
|   Oceania | -0.345 | -3.45 | *** | -0.023 | (-0.036 / -0.010) |
|   Germany | -0.570 | -10.61 | *** | -0.081 | (-0.096 / -0.066) |
|   Netherlands | -0.314 | -6.06 | *** | -0.050 | (-0.065 / -0.035) |
|   United Kingdom | -0.028 | -0.79 | | -0.007 | (-0.022 / 0.008) |
| Publisher (Ref. Others) | | | | | |
|   Elsevier | 0.741 | 12.62 | *** | 0.124 | (0.105 / 0.143) |
|   Emerald | 0.155 | 1.46 | | 0.010 | (-0.004 / 0.024) |
|   SAGE | 0.424 | 5.82 | *** | 0.039 | (0.027 / 0.051) |
|   Springer Nature | 0.163 | 2.76 | *** | 0.025 | (0.008 / 0.042) |
|   Taylor & Francis | -0.546 | -9.80 | *** | -0.094 | (-0.111 / -0.077) |
|   Wiley-Blackwell | 0.233 | 3.77 | *** | 0.032 | (0.016 / 0.048) |
| Number of obs. | 19,976 | | | | |
| F(32, 19943) | 170.62 | | | | |
| Prob > F | 0.000 | | | | |
| R-squared | 0.215 | | | | |
| Root MSE | 1.596 | | | | |

Note: *** Significant at 1%; ** Significant at 5%; * Significant at 10%

Starting with the variables of interest, note that the two variables related to the journals are statistically significant at the 1% significance level and have a positive sign. This means that those journals that are OA or publish AIPs have higher CiteScore values, *ceteris paribus*. This confirms the result obtained by Li et al. (2018) also for the CiteScore but with a different methodology for the OA citation advantage. Of these two variables, the one with the highest influence is the one related to publishing AIPs (with a standardized coefficient between 0.12 and 0.15). That is, the most important factor is not being an OA journal but rather making the journal articles themselves immediately accessible.

Of the two explanatory variables related to the publishers, only one is statistically significant at the 10% significance level, specifically the variable related to the percentage of OA journals of the publisher. The positive sign of the estimation shows that the higher the percentage of OA journals the higher the CiteScore.

The control variables included show some interesting features related to their correlation with the CiteScore. Nearly all are statistically significant. Among the variables related to journals, we can see -in a *ceteris paribus* context- that recent incorporation in the Scopus index correlates negatively with CiteScore, as does publishing articles in an increasing number of languages. On the other hand, publishing in English -or even in Chinese- and being indexed in an increasing number of Scopus subject categories or branches, correlates positively with CiteScore. The highest CiteScore corresponds to the Physical Sciences branch (with a standardized coefficient between 0.26 and 0.32) followed by the Health Sciences, Life Sciences, Social Sciences and the General branches.

The control variables related to publishers refer to their country of publication and to some specific publishing houses. First, it is shown that the average CiteScore in all regions is lower than that of the United States, except for the United Kingdom which shows no statistically significant difference compared to the average CiteScore of journals published in the United States. It is also shown that Elsevier, SAGE, Wiley-Blackwell, and Springer Nature have a positive and statistically significant coefficient, which means that their average publication's CiteScore is higher than that of other publishers, *ceteris paribus*. In contrast, Taylor & Francis publications have a negative and statistically significant coefficient, which means that its average publication's CiteScore is lower than that of other publishers, *ceteris paribus*.

## 4. Conclusion

The communication of research findings has benefited greatly from the emergence of the Internet and especially from some publication practices that have their origins in the widespread use of the web. Nowadays, many journals post accepted articles online before they are formally published in an issue (in-press articles). Moreover, a growing number of journals are making their articles available free of charge (gold open access). Both publication practices aim to increase visibility, usage, and citation impact.

The main objective of this paper was to search for evidence to confirm or refute the hypothesis that advancing the publication of in-press articles, or being an OA journal, improves the average impact of a journal. With this aim, the Scopus dataset was statistically analysed and the following conclusions drawn:

- First, the data reveal evidence that shows the highest important relation to citation advantage is advance publication of accepted articles.
- Second, although the average impact of OA journals is slightly lower than that of non-OA journals, the data indicate that this difference is gradually diminishing and tending to disappear.
- Third, the citation advantage of journals that publish articles in press is increasing with time.
- Finally, using a set of control variables related to language, categories, branches, geographical location and publisher, it was found that those journals that are either OA or publish articles in press have higher CiteScore values in a *ceteris paribus* context. Moreover, publishing articles in press relate stronger to the impact factor of the journal than being OA.

Another collateral conclusion of the research supports the well-known axiom related to the language of scientific publication, showing that research that is communicated in English has more impact.

Finally, after controlling for a set of control variables, the highest average impact factor in 2016 was found to be for Elsevier publications.

Our study does have certain limitations. One first limitation is the impact of self-archiving on the NOA journals. This is because we work at journal level. Paywalled journals often allow authors to deposit preprint or postprint versions of the paper in repositories. According to the SHERPA/RoMEO database of publishers' policies on copyright and self-archiving, 81% of publishers formally allow some form of self-archiving. In this way, paywalled journals benefit from their subscriber network and at the same time from the efforts of many authors who may facilitate access via the green route. In order to control for the self-archiving, we suggest further analyses at article level in future.

A second limitation of our study is the data source. It is based on a database with selective coverage. Scopus tends to feature top international journals rather than lower impact or more nationally oriented journals. This is the case in all research areas but is especially true in the Arts & Humanities and Social Sciences categories. In order to control more nationally oriented journals, we suggest further analyses with other data source as Google Scholar in future.

A third limitation is the impact of publishing cost on the OA journals. We do not take into account the influence of article processing charge (APC) costs. The APCs of top-ranked journals are evidently higher than those of lower ranked ones. For this reason, many authors cannot publish in their preferred gold OA journals, especially the top-ranked ones. In order to control for these costs, we suggest further analyses with data sources that include this information in future.

Our results have managerial implications to stakeholders of journals such as publishers, editors, and authors. The early view is beneficial -both for the publisher and the authors-, as by means of an early distribution they may obtain also an increase in the number of citations. In the case of authors, they are increasingly demanding to make their results accessible in the shortest possible time. In the case of publishers, nowadays there are no actual reasons to delay the publication date of an accepted article, just to make it coincide with a set of articles in a particular journal number. This may be an anachronism of the physical distribution era that has no sense in the digital era anymore.

Therefore, AIP publishing is a strategy that journals can employ to increase the impact factor/citations metrics of a journal, as the early view allows journals to formally publish papers which afterwards will 'born' already with citations. In an environment where publishing times are slow and publication processes are counted in months -and sometimes years-, AIP publishing turns to be a good strategy to be more efficient and competitive.

As a final recommendation, publishers should make articles accessible on a continuous basis, once the version accepted by the editor is available, based on the results of this study, which provides empirical evidence of a citation advantage in this strategy and confirms previous results (Tort, Targino & Amaral, 2012; Yu, Wang & Yu, 2005). In fact, many journals 'born' as open access (e.g., Plos One, Scientific Reports, Nature Communications, Science Advances, SAGE Open) employ this system of continuous publication outside the rigid structure in fixed date issues. These journals are published continuously, so they do not have queues of documents to be published, which significantly reduces the distribution times.